\documentclass[%
 reprint,
bibnotes,
 amsmath,amssymb,
 aps,
floatfix,
]{revtex4-2}

\usepackage{graphicx}% Include figure files
\usepackage{braket}
\usepackage{xcolor}
\usepackage{dcolumn}% Align table columns on decimal point
\usepackage{bm}% bold math
\usepackage{array}
\usepackage{booktabs}
\usepackage{graphicx}
\usepackage{physics}
\usepackage{algorithm2e}

\usepackage{lipsum}

\begin{document}

\preprint{APS/123-QED}

\title{Deep-Microscopic Photonic Bandstructure of a Two-Dimensional Honeycomb Lattice}% Force line breaks with \\
% \thanks{A footnote to the article title}%

\author{Pronoy Das$^{1}$}

\author{Sathwik Bharadwaj$^{2}$}

\author{Wenbo Sun$^{1}$}

\author{Zubin Jacob$^{1}$}%
\email{zjacob@purdue.edu}

\affiliation{
  $^{1}$Elmore Family School of Electrical and Computer Engineering, Birck Nanotechnology Center, and Purdue Quantum Science and Engineering Institute, Purdue University, West Lafayette, Indiana 47907, USA \\
  $^2$Department of Physics, Worcester Polytechnic Institute, Worcester, Massachusetts 01609, USA
}

\date{\today}% It is always \today, today,
%  but any date may be explicitly specified

\begin{abstract}
  Conventional descriptions of light in materials rely on macroscopic models that average the optical response over the microscopic lattice structures. These models are deployed in the semi-classical limit, where the collective local response over many lattice sites is adequate. However, they fail to capture the full spatial dispersion, which requires a lattice-resolved description of light. We address this limitation by formulating a deep-microscopic nonlocal theory of the photon based on the Maxwell Hamiltonian and its characteristic equation. The solution reveals the full spatial dispersion in the form of deep-microscopic photonic bandstructure, and corresponding real-space Bloch field textures. We demonstrate our theory for a two-dimensional (2D) tight-binding model of a gapped honeycomb lattice. Our framework thus extends photons to the same microscopic footing as electrons or phonons, providing a unified route to studying topology, dissipation, and quantum fluctuations of light in dispersive materials.
\end{abstract}

\maketitle

% =========================================================
% I. OVERVIEW
% =========================================================
\section{Introduction}
\label{sec:overview}

\begin{figure*}[!htbp]
  \centering
  \includegraphics[width=0.7\textwidth]{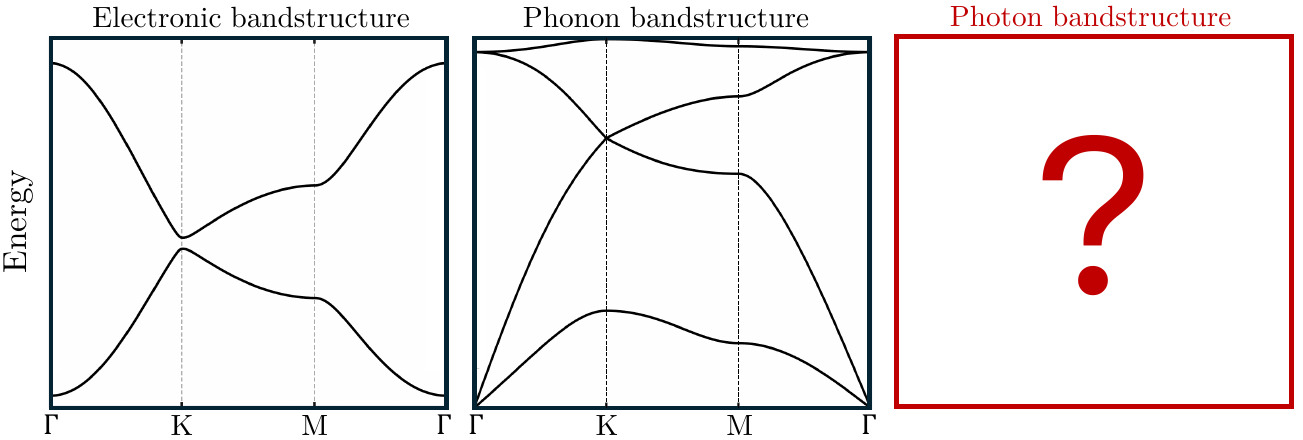}
  \caption{Schematic comparison of microscopic band theories for the three fundamental excitations of the 2D gapped honeycomb lattice used in this work. Electronic and phonon bandstructures are routinely obtained from lattice Hamiltonians that encode the full crystal symmetry. However, a comparable microscopic description for photons has been absent.}
  \label{fig:vision}
\end{figure*}

Microscopic behavior of electrons and phonons in crystalline materials is well described by Hamiltonian-based nonlocal Bloch frameworks that explicitly encode the lattice symmetries \cite{AshcroftMermin,Kittel,BornHuang}. However, no comparable microscopic description exists for the photon. The existing theories remain anchored to macroscopic approximations \cite{Sipe2020}, in which the material is represented by a local (or at most, a weakly nonlocal) optical response \cite{AgranovichGinzburg}. Such macroscopic models have proved remarkably successful for bulk optics, plasmonics, and even nanophotonics \cite{Goncalves2020}. Yet they average the optical response over the lattice sites, leaving them largely agnostic to lattice symmetries. As a result, they fail to capture the intrinsic spatial dispersion of the material \cite{Das2026,BuschChristensen2024}.

A partial remedy exists in the literature for photonic crystals, where  the spatial dispersion of light is routinely recast as a bandstructure \cite{Yablonovitch1987,John1987,Joannopoulos2008, BuschChristensen2024,FanRechtsman2022, Khanikaev2024, Ozawa2019}. However, such bandstructures inform only on the interaction of light with the engineered photonic structures and not the spatial dispersion of the materials themselves. These models often approximate the material response in the local macroscopic regime. A full nonlocal lattice-scale photonic band theory of materials remains an open challenge  (Fig.~\ref{fig:vision}) \cite{Bharadwaj2022,deAbajo2025}.

In this Letter, we resolve the full nonlocal spatial dispersion of the photon in a material as a deep-microscopic photonic bandstructure. This treats the photon as a Bloch field, analogous to the field-theoretical microscopic treatment of the electron and the phonon. To achieve this, we introduce the Maxwell Hamiltonian, which encodes the spatial dispersion of the material in a nonlinear eigenproblem \cite{VanMechelen2019}. The nonlocality of the optical response is fixed by the lattice symmetries of the material  \cite{Das2026,Bharadwaj2025}. The solutions of this Maxwell Hamiltonian eigenproblem reveal the bandstructure of the photon in the deep-microscopic regime, defined by the highly-dispersive domain of low frequency and high momentum. As a demonstration, we resolve the deep-microscopic photonic bandstructure for a two-dimensional gapped (Haldane) honeycomb lattice \cite{Haldane1988}, and reveal the Bloch optical modes as electric field textures within the honeycomb lattice. Note that the spatial dispersion features revealed by this bandstructure are not corrections to a local description. These features are qualitatively absent from any local or weakly nonlocal theory of light in solids.

The remainder of this manuscript details the theoretical formulation and numerical results of this approach. Section~\ref{sec:maxwell_ham} derives the Maxwell Hamiltonian eigenproblem for a general 2D periodic lattice, and then introduces the gapped honeycomb model used as our demonstration platform. Section~\ref{sec:bandstructure} solves this eigenproblem and presents the resulting deep-microscopic photonic bandstructure together with the associated real-space Bloch modes. Finally, Section~\ref{sec:nonlocal_optical_response} details the nonlocal optical response tensor, computed via the extended Adler-Wiser formalism, and shows that the nonlocal light-matter interaction is essential to the band formation.

% =========================================================
% II. MAXWELL HAMILTONIAN EIGENPROBLEM
% =========================================================
\section{The Maxwell Hamiltonian Eigenproblem}
\label{sec:maxwell_ham}

Here, we derive the Maxwell Hamiltonian eigenproblem for a general 2D periodic lattice that is linear, time-invariant, and source-free. In the temporal gauge (scalar potential $\phi=0$), the electromagnetic field separates into in-plane electric components $(E_x,E_y)$ and an out-of-plane magnetic component $H_z$ \cite{Boito2020}. We combine these into the Riemann-Silberstein field vector $\vec{\alpha} = \mathbf{E} + i\mathbf{H}$ \cite{Silberstein1907} and expand in a Bloch-Fourier basis over crystal momenta $\mathbf{q} \in 1^{st}$ BZ (Brillouin Zone) and reciprocal lattice vectors $\mathbf{G}$. Maxwell's equations in a 2D periodic lattice then reduce to the eigenproblem (\cite{Supp})
\begin{equation}
  \mathcal{H}_{M}(\mathbf{q}+\mathbf{G})\vec{\alpha}_{\mathbf{G}}(\mathbf{q},\omega)=\frac{\omega}{c}\sum_{\mathbf{G}'}\mathcal{R}_{\mathbf{G}\mathbf{G}'}(\mathbf{q},\omega)\vec{\alpha}_{\mathbf{G}'}(\mathbf{q},\omega).
  \label{eq:maxwell_ham}
\end{equation}
Here $\vec{\alpha}_{\mathbf{G}}(\mathbf{q},\omega)$ is the reciprocal-space Riemann-Silberstein vector at $\mathbf{G}$, and $\mathcal{H}_{M}$ is the Maxwell Hamiltonian:
\begin{equation}
  \mathcal{H}_{M}(\mathbf{q}')=\mathbf{S}\cdot \mathbf{q}'.
  \label{eq:HM_def}
\end{equation}
Here $\mathbf{S} = (S_x, S_y)$ is the 2D vector of $3\times 3$ spin-1 matrices \cite{VanMechelen2019}:
\[
  S_x = \begin{pmatrix} 0 & 0 & 0 \\ 0 & 0 & -i \\ 0 & i & 0 \end{pmatrix}, \qquad
  S_y = \begin{pmatrix} 0 & 0 & i \\ 0 & 0 & 0 \\ -i & 0 & 0 \end{pmatrix}.
\]
The nonlocal optical response tensor $\mathcal{R}_{\mathbf{G}\mathbf{G}'}(\mathbf{q},\omega)$ couples the Riemann-Silberstein field at reciprocal lattice vector $\mathbf{G}'$ to the material polarization at $\mathbf{G}$. For a nonmagnetic medium (magnetic permeability $\mu=1$):
\begin{equation}
  \mathcal{R}_{\mathbf{G}\mathbf{G}'}(\mathbf{q},\omega)=\left(\begin{array}{ccc}
      \varepsilon_{\mathbf{G}\mathbf{G}'}^{LL} & \varepsilon_{\mathbf{G}\mathbf{G}'}^{LT} & 0                               \\
      \varepsilon_{\mathbf{G}\mathbf{G}'}^{TL} & \varepsilon_{\mathbf{G}\mathbf{G}'}^{TT} & 0                               \\
      0                                        & 0                                        & \delta_{\mathbf{G}\mathbf{G}'}.
    \end{array}\right) \nonumber
  \label{eq:RGG}
\end{equation}
Here $\varepsilon$ is the nonlocal dielectric tensor in the longitudinal/transverse ($L/T$) basis with respect to $\mathbf{q}+\mathbf{G}$. The diagonal magnetic block $\delta_{\mathbf{G}\mathbf{G}'}$ reflects the nonmagnetic condition. We shall discuss further on the microscopic origin and structure of this nonlocal optical response tensor in Section~\ref{sec:nonlocal_optical_response}.

For nonmagnetic media, we eliminate the magnetic field (\cite{Supp}) to obtain a reduced eigenproblem in terms of the electric field alone:
\begin{align}
  \left(|\mathbf{q}+\mathbf{G}|^{2}I-(\mathbf{q}+\mathbf{G})(\mathbf{q}+\mathbf{G})^{T}\right)\mathbf{E}_{\mathbf{G}}(\mathbf{q},\omega) \nonumber \\ =\frac{\omega^{2}}{c^{2}}\sum_{\mathbf{G}'}\varepsilon_{\mathbf{G}\mathbf{G}'}(\mathbf{q},\omega)\mathbf{E}_{\mathbf{G}'}(\mathbf{q},\omega).
  \label{eq:reduced}
\end{align}
The left-hand side is a transverse projection operator that encodes the free-photon kinematics. The right-hand side couples every plane-wave component through the nonlocal dielectric tensor $\varepsilon_{\mathbf{G}\mathbf{G}'}(\mathbf{q},\omega)$.
\begin{figure}[!htbp]
  \centering
  \includegraphics[width=0.8\columnwidth]{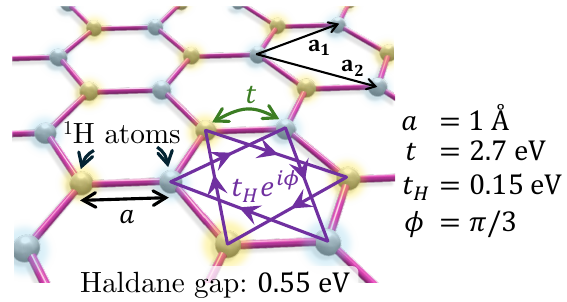}
  \caption{Two-dimensional gapped honeycomb lattice model used as the demonstration platform for the photon eigenproblem. Nearest-neighbor hoppings $t$ (green arrows) connect the two sublattices, while complex next-nearest-neighbor Haldane hoppings $t_{\mathrm{H}}\mathrm{e}^{\mathrm{i}\phi}$ (purple arrows) break time-reversal symmetry and open a gap in the electronic bandstructure (Fig.~\ref{fig:vision}).}
  \label{fig:honeycomb_bands}
\end{figure}

As a demonstration, we adopt the two-dimensional gapped honeycomb lattice with Haldane-type tight-binding couplings \cite{Haldane1988} (Fig.~\ref{fig:honeycomb_bands}). The Maxwell Hamiltonian formalism itself, however, is general and applies to any periodic lattice.

% =========================================================
% III. PHOTON BANDSTRUCTURE
% =========================================================
\section{Deep-Microscopic Photonic Bandstructure}
\label{sec:bandstructure}

The frequency dependence of $\varepsilon_{\mathbf{G}\mathbf{G}'}(\mathbf{q},\omega)$ renders the eigenproblem in Eq.~(\ref{eq:reduced}) nonlinear in $\omega$. We solve this by approximating each element of the dielectric tensor with a sum of Lorentz oscillators:
\begin{equation}
  \varepsilon_{\mathbf{G}\mathbf{G}'}(\omega) \approx \delta_{\mathbf{G}\mathbf{G}'} + C_{\mathbf{G}\mathbf{G}'} + \sum_{p=1}^{N}\frac{S_{\mathbf{G}\mathbf{G}'}^{(p)}}{\omega_{p,\mathbf{G}\mathbf{G}'}^{2} - \omega^{2} - \mathrm{i}\gamma_{\mathbf{G}\mathbf{G}'}^{(p)}\omega},
  \label{eq:eps_lorentz}
\end{equation}
where $C_{\mathbf{G}\mathbf{G}'}$ is a frequency-independent background correction and, for each pole $p$, $S_{\mathbf{G}\mathbf{G}'}^{(p)}$, $\omega_{p,\mathbf{G}\mathbf{G}'}$, and $\gamma_{\mathbf{G}\mathbf{G}'}^{(p)}$ denote the oscillator strength, resonance frequency, and damping rate, respectively. We determine these parameters by least-squares fitting of the full nonlocal $\varepsilon_{\mathbf{G}\mathbf{G}'}(\mathbf{q},\omega)$ over a chosen frequency window. Two Lorentzians ($N=2$) provide an accurate fit over the window $0-6$~eV. Extending the description to higher frequencies requires additional poles at increased computational cost. Substituting Eq.~\ref{eq:eps_lorentz} into Eq.~\ref{eq:reduced} and introducing the change of variable $s = -\mathrm{i}\omega$, the two rational denominators combine into a sixth-order polynomial in $s$. We then collect six auxiliary fields $(\mathbf{E}_{\mathbf{G}},\,s\mathbf{E}_{\mathbf{G}},\,\ldots,\,s^{5}\mathbf{E}_{\mathbf{G}})$ into an extended state vector $\mathbf{X}_{\mathbf{G}}$, which recasts the eigenproblem as linear:
\begin{equation}
  \sum_{\mathbf{G}'} A_{\mathbf{G}\mathbf{G}'}(\mathbf{q})\,\mathbf{X}_{\mathbf{G}'} = s \sum_{\mathbf{G}'} B_{\mathbf{G}\mathbf{G}'}\,\mathbf{X}_{\mathbf{G}'},
  \label{eq:gen_eig}
\end{equation}
where $A_{\mathbf{G}\mathbf{G}'}(\mathbf{q})$ collects the wavevector-dependent kinetic term together with the Lorentz parameters (resonance frequencies, damping rates, oscillator strengths), while $B_{\mathbf{G}\mathbf{G}'}$ is a structural shift matrix carrying only the background contribution $\delta_{\mathbf{G}\mathbf{G}'}+C_{\mathbf{G}\mathbf{G}'}$ (the full expression of the two matrices $A_{\vb{G}\vb{G}^\prime}$ and $B_{\vb{G}\vb{G}^\prime}$ is derived in \cite{Supp}). The total matrix dimension is $6N_{G}\times 6N_{G}$. We recover the physical eigenfrequencies as $\omega(\mathbf{q}) = \mathrm{i}s(\mathbf{q})$. Their real parts yield the photon dispersion, and their imaginary parts quantify photon attenuation.

\begin{figure}[!htbp]
  \centering
  \includegraphics[width=0.9\columnwidth]{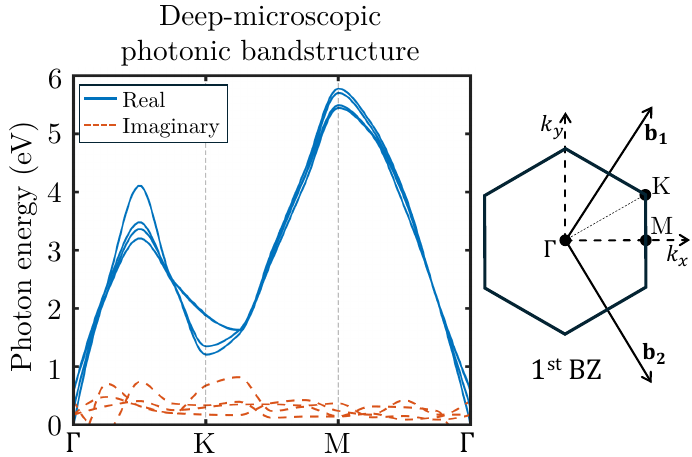}
  \caption{Deep-microscopic photonic bandstructure along high-symmetry directions of the first Brillouin zone, obtained by solving the Maxwell Hamiltonian eigenproblem with the full nonlocal susceptibility tensor derived from the Haldane lattice of Fig. \ref{fig:honeycomb_bands}. Real and imaginary parts of the photon eigenfrequencies are shown, revealing a photonic gap that inherits the topology of the underlying electronic Haldane gap.}
  \label{fig:ph_band}
\end{figure}

Figure~\ref{fig:ph_band} presents the resulting photon bandstructure along the high-symmetry path $\mathbf\Gamma\textrm{-}\mathbf K\textrm{-}\mathbf M\textrm{-}\mathbf \Gamma$ of the $1^{st}$ BZ. For each band $n$ at crystal momentum $\mathbf{q}$, the eigenvector furnishes the Bloch-Fourier coefficients of the electric field. Retaining $N_G=6$ reciprocal-lattice vectors, it takes the form $\mathbf{E}_{\mathbf{G}}(\mathbf{q},\omega_{n})=\left( E_{\mathbf{G}_{0}}^{L}, \ldots, E_{\mathbf{G}_{6}}^{L}, E_{\mathbf{G}_{0}}^{T}, \ldots, E_{\mathbf{G}_{6}}^{T} \right)^T$, where $E_{\mathbf{G}_i}^L$ and $E_{\mathbf{G}_i}^T$ are the longitudinal and transverse polarization components at reciprocal lattice vector $\mathbf{G}_i$. Here $\mathbf G_0=0$ denotes the first Brillouin zone and $\mathbf G_1, \ldots, \mathbf G_6$ the nearest shells.

\begin{figure}[!htbp]
  \centering
  \includegraphics[width=1.0\columnwidth]{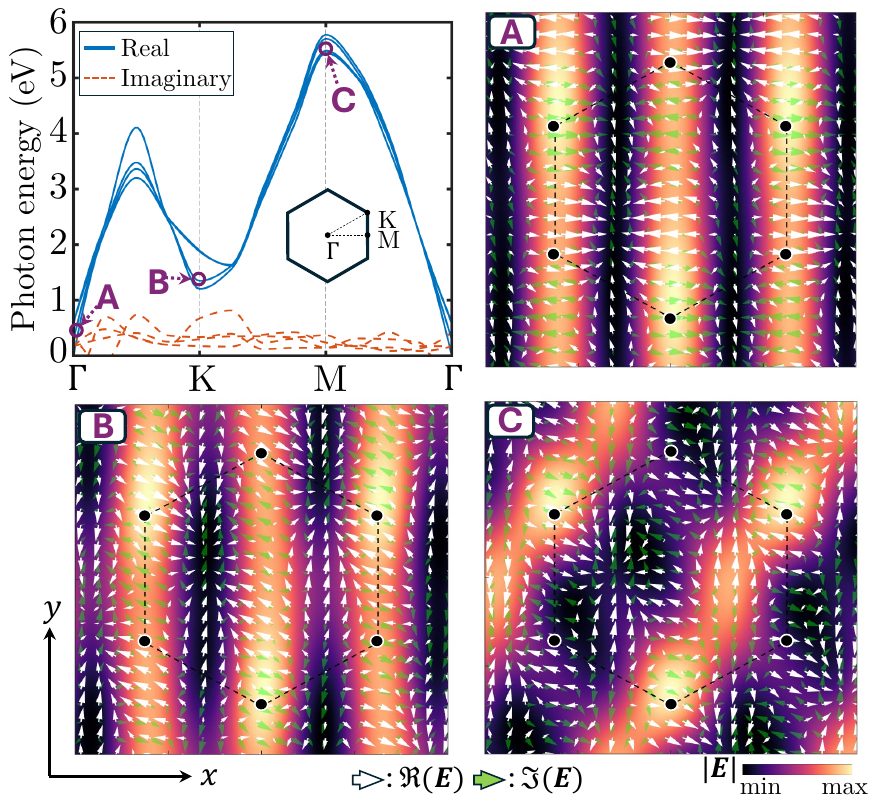}
  \caption{Real-space Bloch electric field textures of selected photon eigenmodes. The color map indicates the field magnitude and the white(green) arrows denote the real(imaginary) components of the polarization vector, both superimposed on the honeycomb lattice (black lines).}
  \label{fig:efield}
\end{figure}

We reconstruct the real-space field profile at lattice site $\mathbf{R}$ from these coefficients as
\begin{equation}
  \mathbf{E}_{\mathbf{R}}(\mathbf{r},t;\mathbf{q},\omega_{n}) = e^{i\mathbf{q}\cdot\mathbf{r}}e^{-i\omega_{n}t}\mathbf{u}(\mathbf{r})
  \label{eq:bloch_field}
\end{equation}
\noindent where $\mathbf{u}(\mathbf{r}) = \sum_{\mathbf{G}} \mathbf{E}_{\mathbf{G}}(\mathbf{q},\omega_n)\, e^{i\mathbf{G}\cdot\mathbf{r}}$ is the lattice-periodic Bloch function and the total field is $\mathbf{E}(\mathbf{r},t;\mathbf{q},\omega_{n})=\sum_{\mathbf{R}}\mathbf{E}_{\mathbf{R}}(\mathbf{r},t;\mathbf{q},\omega_{n})e^{i\mathbf{q}\cdot\mathbf{R}}$. This is the direct photonic analogue of the electronic Bloch wavefunction: the eigenmode factorizes into a plane-wave envelope carrying crystal momentum $\mathbf{q}$ and a periodic function $\mathbf{u}(\mathbf{r})$ that encodes the microscopic lattice symmetry.

Figure~\ref{fig:efield} displays real-space Bloch electric field textures $\mathbf{E}_{\mathbf{R}}(\mathbf{r},t;\mathbf{q},\omega_{n})$ reconstructed via Eq.~\ref{eq:bloch_field}. The lattice strongly modulates these field textures, a structure entirely absent in local or macroscopic descriptions. Such modulation shows that a deep-microscopic framework is required for capturing the full spatial character of light in periodic media.

% =========================================================
% IV. NONLOCAL OPTICAL RESPONSE
% =========================================================
\section{Nonlocal Optical Response}
\label{sec:nonlocal_optical_response}

The Maxwell eigenproblem derived above is universal, and all material-specific information enters through the nonlocal dielectric tensor $\varepsilon_{\mathbf{G}\mathbf{G}'}(\mathbf{q},\omega)$. For the Haldane honeycomb model, we compute this tensor via the extended Adler-Wiser formalism \cite{Adler1962,Wiser1963}. This formalism evaluates the full current-current correlation function at finite photon momentum $\mathbf{q}$ and for all reciprocal-lattice shell pairs $\mathbf{G},\mathbf{G}'$ \cite{Das2026,Bharadwaj2025,Supp}. The higher-order elements $\mathbf{G},\mathbf{G}'\neq 0$ carry the nonlocality. For the off-diagonal elements $\mathbf{G}\neq\mathbf{G}'$, each component couples a field at momentum $\mathbf{q}+\mathbf{G}'$ to the polarization response at $\mathbf{q}+\mathbf{G}$, transferring a reciprocal-lattice momentum $\mathbf{G}-\mathbf{G}'$ to the crystal. This is directly analogous to electronic Umklapp scattering. Conventional local macroscopic optics retains only the $\mathbf{G}=\mathbf{G}'=0$ component \cite{Bharadwaj2025,Das2026}.

\begin{figure}[!htbp]
  \centering
  \includegraphics[width=1.0\columnwidth]{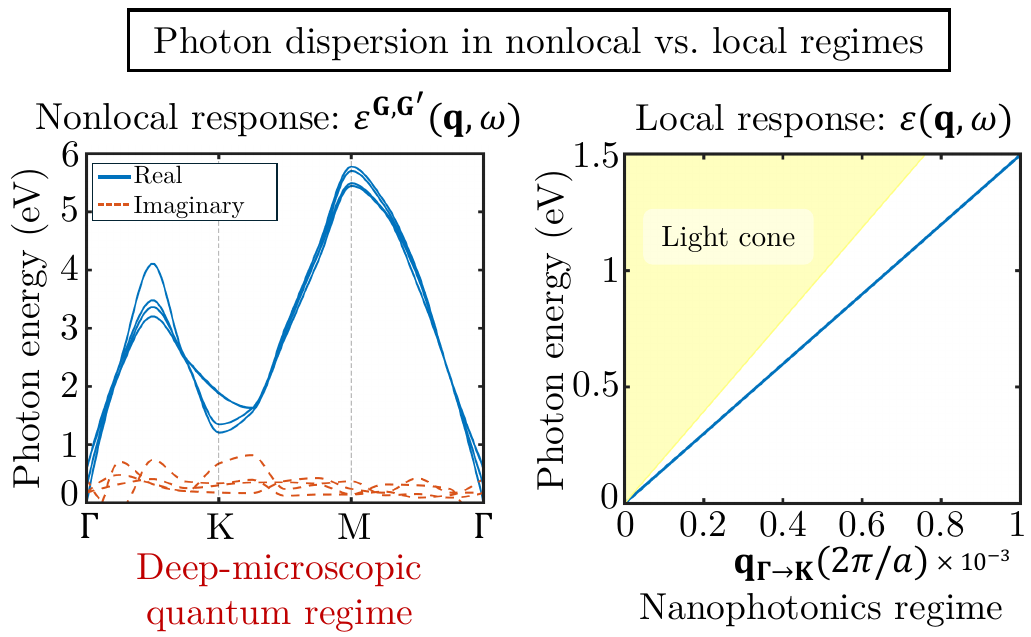}
  \caption{Comparison between the photon dispersion obtained with the full nonlocal optical response and the local approximation ($\mathbf{G}=\mathbf{G^\prime}= 0$). The local approximation fails to capture the deep-microscopic photonic bands within the lattice.}
  \label{fig:comp}
\end{figure}

\begin{figure}[!htbp]
  \centering
  \includegraphics[width=0.5\columnwidth]{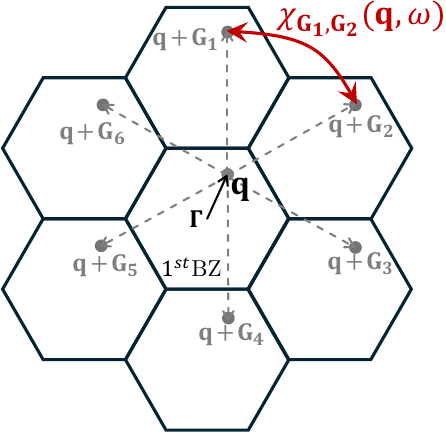}
  \caption{Structure of the nonlocal susceptibility tensor in reciprocal space. A field perturbation at momentum $\mathbf{q}+\mathbf{G}_m$ couples to the response at all $\mathbf{q}+\mathbf{G}_n$ through the tensor $\chi_{\mathbf{G}_m\mathbf{G}_n}(\mathbf{q},\omega)$, which enters the Maxwell Hamiltonian as a dense matrix mixing every plane-wave component. The off-diagonal elements, absent in local theories, encode the lattice-scale spatial dispersion that gives rise to the photon bandstructure.}
  \label{fig:nonlocal_chi}
\end{figure}
The deep-microscopic photon bands presented in this work emerge from nonlocal light-matter interactions. Retaining only the diagonal $\mathbf{G}=\mathbf{G}'=0$ component recovers the conventional light cone but fails to generate the lattice-scale band structure (Fig.~\ref{fig:comp}). Only by including the off-diagonal inter-shell couplings does the photon dispersion acquire the folding, gapping, and hybridization characteristic of a true Bloch band theory (Fig.~\ref{fig:nonlocal_chi}). The nonlocal optical response is therefore not a perturbative correction. It is the essential ingredient that elevates photonic dispersion from a featureless continuum to a rich bandstructure governed by the lattice symmetries.
% =========================================================
% VI. CONCLUSIONS
% =========================================================
\section{Conclusion}
\label{sec:conclusion}

We have formulated a Maxwell Hamiltonian that yields the deep-microscopic photonic bandstructure for a general 2D periodic lattice. The light-matter interaction enters through the full nonlocal optical response tensor, which we compute via the extended Adler-Wiser formalism. For a honeycomb lattice with Haldane-type gap and tight-binding coupling, the resulting photon bands exhibit gaps inherited from the electronic topology \cite{Ozawa2019}, complex eigenfrequencies that encode damping losses, and real-space Bloch electric field textures modulated by the lattice symmetry. The local-response limit fails to reproduce these features, confirming that off-diagonal inter-shell couplings are essential to the bandstructure.

Probing the deep-microscopic photon dispersion experimentally requires sub-lattice spatial resolution and large momentum transfer, which cannot be achieved with standard far-field optics \cite{AgranovichGinzburg}. Momentum-resolved THz spectroscopy offers a direct route to mapping the large-wavevector photon dispersion. In addition, ultra-nearfield scanning optical microscopy can resolve the lattice-modulateds Bloch fields \cite{Cocker2021}.

\section{Acknowledgements}
\label{sec:Acknowledgements}
This work was supported by the Office of Naval Research (ONR) under the award number N00014231270.

\section{Data Availability}
The data that support the findings of this study are available from the corresponding author upon reasonable request.

\bibliographystyle{plain}
\bibliography{biblio}

\end{document}